\documentclass{ieeeaccess}
\usepackage{hyperref}
\usepackage{pifont}
\usepackage{lmodern}
\usepackage{amsmath,amssymb,amsfonts}
\usepackage{graphicx}
\usepackage{multirow,multicol}
\usepackage{algorithm,algorithmic}
\usepackage{subcaption}
\usepackage{booktabs}
\usepackage{color,soul}

\soulregister\cite7
\soulregister\ref7
\soulregister\pageref7

\def\BibTeX{{\rm B\kern-.05em{\sc i\kern-.025em b}\kern-.08em
 T\kern-.1667em\lower.7ex\hbox{E}\kern-.125emX}}
\begin{document}
\history{Date of publication xxxx 00, 0000, date of current version xxxx 00, 0000.}
\doi{10.1109/ACCESS.2017.DOI}

\title{An Adaptive Scoring Framework for Attention Assessment in NDD Children via Serious Games}

\author{
\uppercase{Abdul Rehman}\authorrefmark{1},
\uppercase{Ilona Heldal}\authorrefmark{1},
\uppercase{Cristina Costescu}\authorrefmark{2},
\uppercase{Carmen David}\authorrefmark{2},
\uppercase{Jerry Chun-Wei Lin}\authorrefmark{1}
}

\address[1]{Department of Computer Science, Electrical Engineering and
Mathematical Sciences, Western Norway University of Applied Sciences,
Bergen, Norway}
\address[2]{Special Education Department, Babeș-Bolyai University, Cluj-Napoca, Romania}

\markboth
{Author \headeretal: Preparation of Papers for IEEE TRANSACTIONS and JOURNALS}
{Author \headeretal: Preparation of Papers for IEEE TRANSACTIONS and JOURNALS}

\corresp{Corresponding author: arj@hvl.no, jerry.chun-wei.lin@hvl.no}

\tfootnote{}
\begin{abstract}
This paper introduces an innovative adaptive scoring framework for children with Neurodevelopmental Disorders (NDD) that is attributed to the integration of multiple metrics, such as spatial attention patterns, temporal engagement, and game performance data, to create a comprehensive assessment of learning that goes beyond traditional game scoring. The framework employs a progressive difficulty adaptation method, which focuses on specific stimuli for each level and adjusts weights dynamically to accommodate increasing cognitive load and learning complexity. Additionally, it includes capabilities for temporal analysis, such as detecting engagement periods, providing rewards for sustained attention, and implementing an adaptive multiplier framework based on performance levels. To avoid over-rewarding high performers while maximizing improvement potential for students who are struggling, the designed framework features an adaptive temporal impact framework that adjusts performance scales accordingly. We also established a multi-metric validation framework using Mean Absolute Error (MAE), Root Mean Square Error (RMSE), Pearson correlation, and Spearman correlation, along with defined quality thresholds for assessing deployment readiness in educational settings. This research bridges the gap between technical eye-tracking metrics and educational insights by explicitly mapping attention patterns to learning behaviors, enabling actionable pedagogical interventions.
\end{abstract}
\titlepgskip=-15pt
\maketitle 
\IEEEpeerreviewmaketitle

\section{Introduction and Related Work}\label{intro}
According to the Diagnostic and Statistical Manual of Mental Disorders, Fifth Edition, Text Revision (DSM-5-TR), neurodevelopmental disorders (NDD) are a group of conditions with onset in the developmental period, typically manifesting early in development, often before a child enters grade school\cite{edition1980diagnostic}. They are characterized by developmental deficits that produce impairments of personal, social, academic, or occupational functioning \cite{edition1980diagnostic}. These deficits can range from particular limitations in learning or control of executive functions to global impairments of social skills or intelligence \cite{edition1980diagnostic}. These include but are not limited to autism spectrum disorder (ASD), attention-deficit/hyperactivity disorder (ADHD), and dyslexia, and put forward individual challenges for children suffering from them \cite{von2022neurodevelopmental}. Tackling those issues is crucial, as it enhances children’s academic achievement and positively impacts the quality of life for children with NDD. One of the tools used by researchers is eye-tracking technology, which is operated under the concept of identifying and interpreting characteristics of eye movements and the visual point of focus to assess attention and cognitive functions \cite{skaramagkas2021review}. Eye tracking can be used to increase the understanding of students’ activities while playing serious games. Serious games are specially designed games intended for purposes beyond pure entertainment, such as education, training, or behavior modification. Eye scans provide information on which objects are most effectively attracting learners’ attention and the areas that appear to escape it \cite{coskun2022systematic}. Based on this information, it becomes easier to match the strategies used with the learner, thus accelerating the learning process. General education offers many opportunities in terms of eye tracking. It can be used in creating educational stimuli that are more easily understood by the target audience, in early diagnosis of learning problems, and in determining the effectiveness of the employed teaching strategies \cite{sharma2022student}. 

\subsection{Related Work Limitations}

Recent research has increasingly explored the intersection of eye-tracking technology and digital interventions for neurodevelopmental disorders. Garcia-Baos et al. \cite{garcia2019novel} developed an interactive eye-tracking game for ADHD attention training, demonstrating significant improvements in impulsivity, reaction time, and fixation control after three weeks of intervention. Merzon et al. \cite{merzon2022eye} utilized virtual reality environments with 90 Hz eye tracking to detect ADHD patterns in children through lifelike prospective memory games. In contrast, Aminoleslami et al. \cite{aminoleslami2023classification} focused on classification approaches using eye tracking data from computer games to distinguish between autistic and typically developing children. Advanced diagnostic frameworks \cite{oliveira2021computer} have emerged, such as computer-aided autism diagnosis systems that utilize visual attention models with genetic algorithm-based feature selection, and comprehensive eye-tracking systems \cite{lee2023use} for ADHD identification, which combine continuous performance tests with gaze analysis.

\subsection{Motivation}

The existing approaches primarily focus on either therapeutic training applications or binary classification tasks using basic eye-tracking metrics. The primary motivation for developing this eye-tracking assessment framework is to move beyond the limitations of traditional correct/incorrect scoring systems, which only indicate whether students provided the correct answer but offer no insight into how they learned or why they struggled, as shown in the game developed in \cite{costescu2023mushroom}. This new framework provides a detailed analysis of the learning process, enabling us to determine whether students are attending to the relevant information, how efficiently they process it, whether they are easily distracted, and how well they can maintain focus over extended periods. Instead of waiting until students fail a test to know they are struggling, teachers can identify attention problems early and provide targeted help. For example, two students might both score 70\% on a test. Still, the eye-tracking data might reveal that one student has excellent focus but lacks knowledge of the stimulus (and needs academic support). At the same time, the other knows the material but has difficulties with attention regulation (and requires cognitive training). This approach transforms education from simply measuring what students got wrong to understanding why they are struggling and how to help them learn more effectively, making assessment a tool for improvement rather than just evaluation. Therefore, this paper significantly contributes to the field of educational eye-tracking and learning analytics:


\subsection{Contributions}
\begin{itemize}
 \item \textbf{
Multi-Dimensional level Adaptive Scoring Framework}: Designed and developed a novel adaptive scoring framework that integrates several metrics such as  (1) spatial attention patterns, (2) temporal engagement and (3) game performance data to provide holistic learning assessment beyond traditional game scoring. It further follows a progressive difficulty adaptation method (from 30\% to 50\% to 65\% content focus targets) with level-specific stimulus focus targets and dynamic weight adjustments that account for increasing cognitive load and learning complexity.
 
 \item \textbf{Advanced Temporal Enhancement feedback Layer}: Integrated temporal analysis capabilities, including engagement period detection (400ms minimum), sustained attention rewards (2500ms threshold), and adaptive multiplier frameworks based on performance levels. Furthermore, it implemented an adaptive temporal impact system that provides performance-scaled adjustments, preventing over-rewarding of high performers while maximizing improvement potential for struggling students.

 \item \textbf{Validation Method}: Established a multi-metric validation framework using MAE, RMSE, Pearson, and Spearman correlations with defined quality thresholds for deployment readiness assessment in educational settings.
 
 \item \textbf{Educational Interpretation Framework}: Bridged the gap between technical eye-tracking metrics and educational insights by providing explicit mappings between attention patterns and learning behaviors, enabling actionable pedagogical interventions.
\end{itemize}

The paper is organized as follows. The following section \ref{scoringmodel} provides the proposed scoring model and its formation. Section \ref{experimentalfinding} provides the findings, and finally, Section \ref{conclusion} concludes the paper.

\section{Scoring Model and Mathematical Formulation}\label{scoringmodel}
\begin{figure*}[!ht]
\centering
\includegraphics[width=\textwidth]{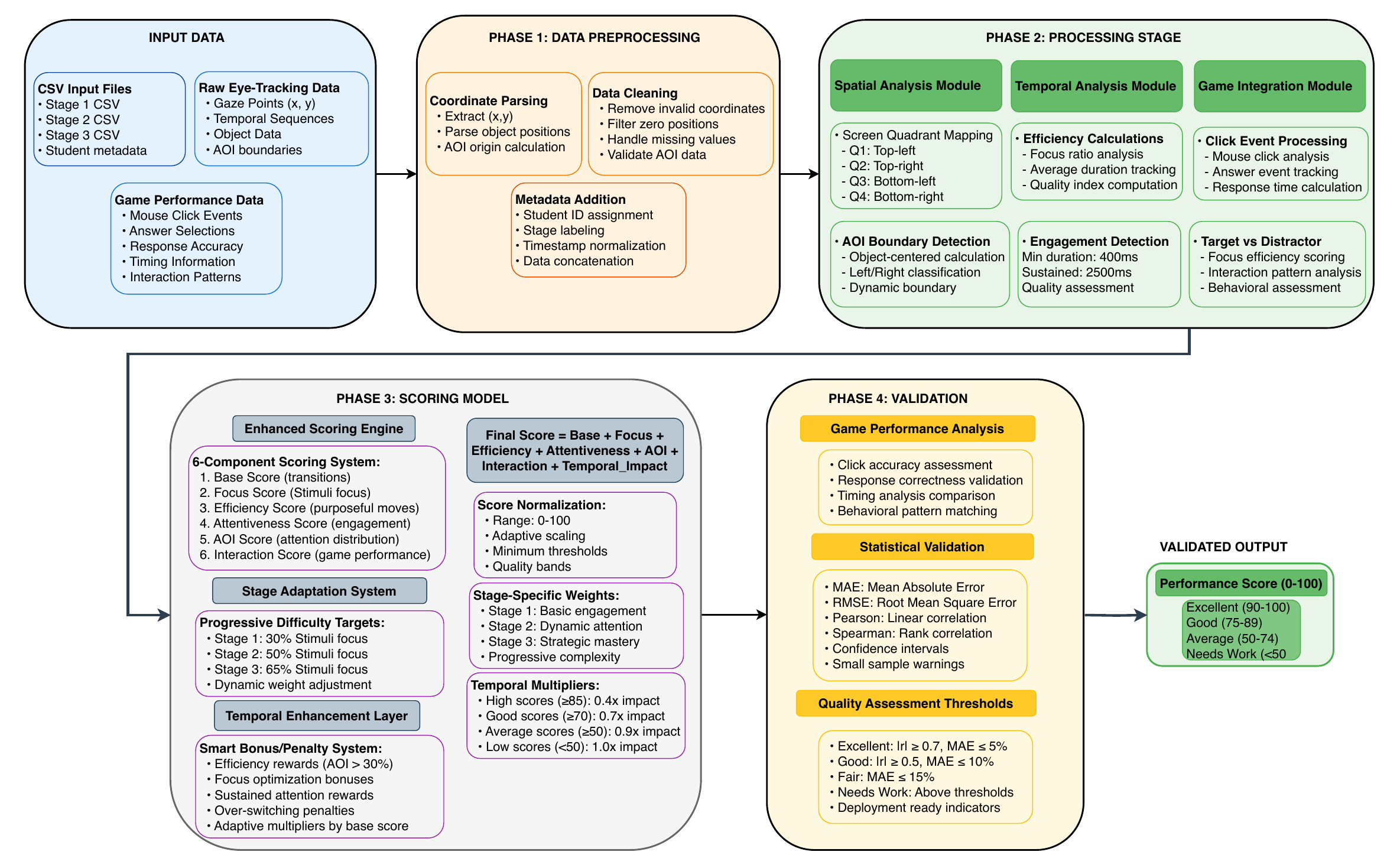}
\caption{Enhanced Eye-Tracking Analysis model Flow}
\label{modelflow}
\end{figure*}

Our proposed model employs a systematic four-phase approach that transforms raw eye-tracking data into meaningful educational assessments, as illustrated in FIGURE \ref{modelflow}. Initially, the model begins with input data in CSV files containing raw eye gaze data, along with temporal features and other game performance events \cite{costescu2023mushroom}. Next, the model pre-processes this multi-level CSV dataset, parsing gaze coordinates and object positions and filtering out invalid data points. It then applies our object-centered AoI classification algorithm to accurately determine when students focus on a specific object, eliminating spatial shift errors present in traditional systems. It combines spatial attention analysis, temporal engagement patterns, and game performance validation into a unified scoring model. The model performs quadrant-based attention analysis, dividing the screen into menu and stimulus areas to track navigation patterns and engagement levels. Subsequently, the algorithm adapts to different difficulty levels by applying level-specific weights and thresholds, ultimately producing a validated performance score that correlates with actual game outcomes while providing detailed insights into visual attention and cognitive engagement patterns. Finally, the system validates performance through Mean Absolute Error (MAE), Root Mean Square Error (RMSE), Pearson correlation, and Spearman correlation. It generates comprehensive visualizations, including AoI validation plots, learning trajectory analysis, and automated performance classification into Mastery, Developing, or Struggling categories, enabling real-time educational intervention and personalized learning support.

\subsection{Game Description}

The sustained attention game used for data collection was formally structured and designed by Costescu et al.~\cite{costescu2023mushroom}. It follows the original task design developed by Rosvold et al.~\cite{CCPT} as an implementation of the Computerized Continuous Performance Task (CCPT). Participants are shown a road in the forest, where mushrooms (\textit{targets}) and flowers(\textit{distractors}) appear randomly on either side. The objective is to touch the screen when a target appears while avoiding interaction with distractors. A Tobii Pro Nano eye-tracker \footnote{https://connect.tobii.com/s/article/how-to-configure-your-tobii-pro-nano-x2-or-x3-eye-tracker-with-the-mobile-device-stand?} was used for data collection, and the game utilizes a custom calibration and head positioning screen explicitly designed for the target demographic. For this paper, we used the Pilot 3 data collected from 10 students who played for 3 levels. Data was collected at a special education school in Romania \cite{costescu2023mushroom}. Each data collection session lasted approximately 1 hour and consisted of three games designed to train different cognitive functions, each with three levels. Our dataset is constructed from data collected for the EU-funded EMPOWER project (grant agreement No. 101060918; see Section \ref{ack} for details), which includes 9 games designed to teach regulation strategies for various cognitive functions. While all games are necessary for the goals of the EMPOWER project~\cite{projectempowerProjectEMPOWER}, eye-tracking data collection features were a priority in all games at this project level. Gaze data were not collected during most games to reduce session length from calibration procedures and maintain a more reliable dataset.

\subsection{Data Pre-Processing}
The first step in the eye-tracking scoring framework is to clean and prepare the raw input data for each difficulty level (levels 1, 2, and 3). However, this raw data is not immediately ready for analysis and requires several pre-processing steps. First, the framework loads CSV files and extracts the eye position coordinates. The raw data stores these positions as text strings, such as "(1250, 680)". Therefore, the pre-processing step converts these strings into numerical values suitable for calculations. The framework also extracts object positions and area-of-interest boundaries in the same way. Next, the framework cleans the data by removing invalid entries. This includes discarding records where the eye tracker could not detect the student's gaze (indicated by zero coordinates) or where tracking was inaccurate. Additionally, any incomplete data that could lead to errors later in the analysis is filtered out. Finally, the pre-processing phase adds the necessary metadata to each record. This includes labeling the student associated with the data, indicating the difficulty level from which it originated, and normalizing the timestamps so that all data points are properly sequenced. The three separate level files are then combined into one comprehensive dataset. After pre-processing, the cleaned dataset contains valid eye-tracking coordinates, game interaction events, and properly labeled metadata, all of which are ready for the in-depth analysis that follows. 

\subsection{Mathematical Model Formulation}
In this section, we present the mathematical formulation of our eye-tracking analysis approach, providing the theoretical foundation for our scoring framework and pattern identification methods. This provides a comprehensive model for transforming raw eye-tracking data into validated performance scores while maintaining interpretability and correlation with actual game outcomes.

Let $\mathcal{D} = \{D_1, D_2, D_3\}$ represent the dataset containing eye-tracking records across three learning levels $l_i \in \{1, 2, 3\}$, where each record $r_i \in D_l$ contains gaze coordinates $GP = \{gp_1, gp_2, \dots, gp_n\}$ be the set of gaze points recorded during an eye-tracking session, object positions $(obj_{x_i}, obj_{y_i})$, AoI dimensions $(w_i, h_i)$, timestamps $t_i$, and game object metadata $O_i$. The screen space is defined as $\Omega = [0, W] \times [0, H]$ where $W = 1920$ and $H = 1080$ pixels. The objective is to construct a performance assessment function $\mathcal{F}: D \rightarrow [0, 100]$ that maps eye-tracking data to a performance score while maintaining a high correlation with actual game performance metrics. The screen is divided into four quadrants $Q = \{Q_1, Q_2, Q_3, Q_4\}$ based on screen center $(w/2, h/2)$ having predefined width $W$ and height $H$ as shown in Eq. \eqref{q1q2a3a4}: 

\begin{align}
\label{q1q2a3a4}
Q(x,y) = \begin{cases}
Q_1 & \text{if } (x,y) \in [0, W/2) \times (H/2, H] \quad \\
Q_2 & \text{if } (x,y) \in [W/2, W] \times (H/2, H] \quad \\
Q_3 & \text{if } (x,y) \in [0, W/2) \times [0, H/2] \quad \\
Q_4 & \text{if } (x,y) \in [W/2, W] \times [0, H/2] \quad 
\end{cases}
\end{align}

Stimulus quadrants $SQ = \{Q_3, Q_4\}$ represent the primary stimuli area, while non-stimulus quadrants $NSQs = \{Q_1, Q_2\}$ represent menu and interface elements.

\textbf{Area of Interest (AoI) Detection: } Each object in level $s\_i$ has a position $(obj_x, obj_y)$ with dimensions $(w_{AoI}, h_{AoI})$. An area of interest boundary rectangle is designed around this object to validate whether the gaze points are on the object or near the object. The AoI boundary is defined as Eq. \eqref{eq2}:

\begin{equation}\label{eq2}
\begin{aligned}
AoI_{obj} = \{(x,y) :\ & obj_x - \frac{w_{AoI}}{2} \leq x \leq obj_x + \frac{w_{AoI}}{2}, \\
 & obj_y - \frac{h_{AoI}}{2} \leq y \leq obj_y + \frac{h_{AoI}}{2} \}
\end{aligned}
\end{equation}

We define two AoIs: one for the object appearing in Q3 and the other in Q4, defined as $\phi: \mathbb{R}^2 \rightarrow \{\text{Left\_AoI}, \text{Right\_AoI}, \text{Outside\_AoI}\}$, defined as Eq. \eqref{eq3}:
{\scriptsize
\begin{align}\label{eq3}
\phi(x,y) = \begin{cases}
\text{Left\_AoI} & \text{if } (x,y) \in AoI_{obj} \text{ and } obj_x < w/2 \\
\text{Right\_AoI} & \text{if } (x,y) \in AoI_{obj} \text{ and } obj_x \geq w/2 \\
\text{Outside\_AoI} & \text{otherwise}
\end{cases}
\end{align}
}

\subsubsection{Transition Analysis Model}

The quadrant transition frequency matrix $T \in \mathbb{R}^{4 \times 4}$ captures movement patterns between quadrants as shown in Eq. \eqref{eq4}:

\begin{align}
\label{eq4}
T_{ij} = |\{k : q(d_k) = Q_i \text{ and } q(d_{k+1}) = Q_j\}|
\end{align}

where $q(d_k)$ returns the quadrant containing gaze point $d_k$. We define transitions between non-stimulus quadrants (NSQs) (Q1 and Q2) and stimulus quadrants (SQs) (Q3 and Q4) as follows in Eqs. \eqref{NSQtostimuli}:

\begin{equation}
\label{NSQtostimuli}
\begin{split}
 T_{NSQ \to SQ} &= |\{t_j \in T | t_j = (Q1 \text{ or } Q2, Q3 \text{ or } Q4)\}| \\
 T_{SQ \to NSQ} &= |\{t_j \in T | t_j = (Q3 \text{ or } Q4, Q1 \text{ or } Q2)\}|\\
 T_{NSQ \to NSQ} &= \left|\left\{ t_j \in T \,\middle|\, t_j = (Q_1, Q_2) \text{ or } t_j = (Q_2, Q_1) \right\}\right|\\
 T_{SQ \to SQ} &= \left|\left\{ t_j \in T \,\middle|\, t_j = (Q_3, Q_4) \text{ or } t_j = (Q_4, Q_3) \right\}\right|
\end{split}
\end{equation}
and total transition made at each S\_i is defined as T\_{total} as shown in Eq. \eqref{eq6}:
\begin{align}
\label{eq6}
T_{total} &= \sum{(T_{NSQ \to SQ}, T_{SQ \to NSQ}, T_{NSQ \to NSQ},T_{SQ \to SQ})}
\end{align}

\textbf{AoI Transition Metrics:}
AoI transitions are quantified by the matrix $A \in \mathbb{R}^{3 \times 3}$ where indices represent: Left\_AoI, Right\_AoI, Outside\_AoI. Key AoI metrics are computed as Eq. \eqref{eq7}, \eqref{eq7a}, and \eqref{eq7b}:

\begin{align}
\label{eq7}
AoI_{Left \leftrightarrow Right} &= AoI_{1 \rightarrow 2} + AoI_{2 \rightarrow 1} \\\label{eq7a}
AoI_{balance} &= \frac{|AoI_{Left} - AoI_{Right}|}{\max(AoI_{Left} + AoI_{Right}, 1)} \\
AoI_{efficiency} &= \frac{AoI_{Left \leftrightarrow Right}}{AoI_{total}}\label{eq7b}
\end{align}

where:\\
- $AoI_{Left} = \sum_{j=1}^{3} A_{1j}$ (total fixations in Left\_AoI)\\
- $AoI_{Right} = \sum_{j=1}^{3} A_{2j}$ (total fixations in Right\_AoI)\\
- $AoI_{total} = \sum_{i,j=1}^{3} A_{ij}$ (total AoI transitions)\\
- $AoI_{1 \rightarrow 2}$ represents transitions from Left\_AoI to Right\_AoI\\
- $AoI_{2 \rightarrow 1}$ represents transitions from Right\_AoI to Left\_AoI\\

\subsubsection{Temporal Engagement Model}
The temporal engagement detection model builds upon established eye-tracking methodologies for identifying meaningful fixation sequences \cite{holmqvist2011eye, duchowski2017eye}. An engagement period $e = (t_{start}, t_{end}, AoI_{type})$ is identified as shown Eq. \eqref{Engagement}:

{\scriptsize
\begin{align}
e \in E \iff 
\begin{cases}
\phi(d_i) = \phi(d_j) \neq \text{Outside\_AoI} 
& \forall i,j \in [k, k+m] \\
t_{k+m} - t_k \geq 400\,\text{ms}
& \text{(Minimum duration)} \\
\nexists\, l \in [k, k+m] : \phi(d_l) \neq \phi(d_k) 
& \text{(Consistency)}
\end{cases}
\label{Engagement}
\end{align}
}
\vspace{4pt}
\noindent
\textbf{Same AoI Condition:} All gaze points must be in the same AoI (Left\_AoI or Right\_AoI), not Outside\_AoI.\\
\textbf{Minimum Duration:} Gaze must stay in that AoI for at least 400ms to filter glances.\\
\textbf{Consistency Check:} Gaze must remain on the same AoI, with no switching during this period.

Here, the minimum threshold varies as per various studies \cite{rayner2009,henderson2003human} (e.g., 200-600ms). Therefore, we select a 400ms minimum threshold based on cognitive processing literature, which shows that meaningful fixations range from 400ms for visual information processing.

\textbf{Sustained Attention Classification:} For sustained attention measurement, we follow established research on cognitive engagement and attention maintenance to select a threshold. We choose a 2500ms threshold that aligns with sustained attention research, which shows optimal engagement periods of 2-3 seconds for complex cognitive tasks \cite{posner2011attention,fan2002testing,maclean2010intensive}. Engagement periods are classified as sustained if their duration exceeds the threshold as shown in Eq. \eqref{eq8}:

\begin{align}
\label{eq8}
E_{sustained} = \{e \in E : duration(e) \geq \tau_{sustained} = 2500ms\}
\end{align}

\textbf{Temporal Efficiency Metrics}
Temporal efficiency calculations are adapted from established eye-tracking metrics for attention allocation and cognitive load assessment \cite{just1976eye} as shown in Eq. \eqref{eqtemporal}, \eqref{eqtemporala}, \eqref{eqtemporalb}:

\begin{align}\label{eqtemporal}
\eta_{temporal} &= \frac{\sum_{e \in E} duration(e)}{session\_duration} \\\label{eqtemporala}
\mu_{engagement} &= \frac{\sum_{e \in E} duration(e)}{|E|} \\\label{eqtemporalb}
\sigma_{sustained} &= \frac{|E_{sustained}|}{|E|}
\end{align}

$\eta_{temporal}$ represents the proportion of total session time spent actively engaging with game objects, calculated by dividing the cumulative duration of all engagement periods by the total session duration. $\mu_{engagement}$ captures the mean duration of individual engagement episodes, computed by dividing the total engagement time by the number of engagement periods. This indicates the typical depth of processing during each focused attention episode, as longer durations generally suggest more thorough cognitive processing of the visual information. $\sigma_{sustained}$ measures the proportion of engagement periods that exceed the sustained attention threshold, calculated by dividing the number of sustained periods by the total number of engagement periods, which reflects the participant's ability to maintain prolonged focus on relevant game elements rather than engaging in brief, superficial glances. These metrics provide quantitative measures of attention distribution and engagement quality, following methodologies established in cognitive workload research \cite{paas2016cognitive, sweller1998cognitive}.

\subsubsection{Enhanced Scoring Function}
The base performance score incorporates transition patterns and focus metrics as in Eq.\eqref{Base} :

{\scriptsize
\begin{align}\label{Base}
S_{base}(s) &= \alpha_1 \cdot T_{SQ} - \alpha_2 \cdot T_{NSQ} + \beta_s \cdot \Gamma_s + \gamma \cdot \Omega_{AoI} + \delta \cdot \Psi_{focus}
\end{align}
}

where $\alpha_1 = 3, \alpha_2 = 1.5$ are transition weights, $\beta_s$ represents level-specific bonuses, $\Gamma_s$ captures level-appropriate transition patterns, $\Omega_{AoI}$ quantifies AoI interaction quality and $\Psi_{focus}$ measures stimuli focus efficiency

\textbf{Level-Specific Adaptations}
The level-specific adaptation in $\beta_l$ reflects the progressive cognitive demands across the three-game difficulty levels. Level 1 emphasizes fundamental object interactions with lower weights, as beginners primarily need to learn fundamental engagement patterns. Level 2 increases the weight on AoI transitions to reward dynamic attention switching between objects, indicating the development of comparison skills. Level 3 places the highest emphasis on bidirectional AoI transitions, reflecting the advanced strategic comparison abilities required at the most challenging level.
The scoring function adapts to level difficulty through parameter modulation as shown in Eq. \eqref{levela}:

\begin{align}
\beta_s = 
\begin{cases}
0.2 \cdot focus_{AoI} + 0.5 \cdot interactions & l = 1 \\
0.25 \cdot focus_{AoI} + 1.8 \cdot transitions_{AoI} & l = 2 \\
0.3 \cdot focus_{AoI} + 2.5 \cdot bidirectional_{AoI} & l = 3
\end{cases}
\label{levela}
\end{align}

\textbf{Stimuli Focus Optimization}
The Stimuli focus (SF) optimization function adjusts scoring thresholds to align with the expected attention patterns for each level of difficulty. Level 1 applies a minimal threshold (25\%) with a modest scoring multiplier (0.4x), recognizing that beginners may spend considerable time in NSQ while learning the interface. The indicator function, denoted as $\mathbb{I}$(condition), serves as a mathematical "switch" that returns 1 when a specified condition is true and 0 when false, enabling precise threshold-based scoring. Level 2 increases both the threshold (40\%) and scoring weight (0.6x), with an additional bonus for optimal focus ranges (50-75\%), reflecting intermediate players' improved ability to concentrate on game Stimuli. Level 3 implements the highest threshold (50\%) and maximum scoring weight (0.75x), plus a substantial bonus for sustained Stimuli focus above 65\%, which corresponds to the advanced attention control expected at the most challenging difficulty level.

The Stimuli focus percentage is computed and optimized per level as shown in Eq. \eqref{eq:stimuli_focus}, \eqref{eq:focus_scoring}:

{\scriptsize
\begin{align}
SF_s &= \frac{\text{time}(Q_3) + \text{time}(Q_4)}{\text{total\_session\_time}} \times 100 \label{eq:stimuli_focus} \\[6pt]
\Psi_{\text{focus}} &=
\begin{cases}
\max(0, -(25 - SF_s) \times 0.4) & s = 1 \\[4pt]
\max(0, -(40 - SF_s) \times 0.6) + \mathbb{I}(50 < SF_s < 75) \times 7.5 & s = 2 \\[4pt]
\max(0, -(50 - SF_s) \times 0.75) + \mathbb{I}(SF_s > 65) \times 10 & s = 3
\end{cases}
\label{eq:focus_scoring}
\end{align}
}

\subsubsection{Temporal Enhancement Integration}
The temporal enhancement function integrates multiple engagement quality measures to modify the base performance score through a framework of bonuses and adjustments. The AoI bonus function rewards efficient attention allocation by providing up to 6 points for temporal efficiency above 30\% while adjusting for limited focus below 10\% with up to 4 points reduced. The sustained attention bonus awards up to 4 points based on the number of prolonged engagement episodes, encouraging deep processing over superficial scanning. We have found that when participants engage in more than 8 different periods, their overall performance tends to be impacted. This usually happens when attention gets spread across too many areas, making it harder to maintain the focused approach that leads to the best outcomes. Our scoring reflects this pattern to help guide more effective engagement strategies.

The temporal enhancement function $\mathcal{T}: \mathbb{R} \times \mathcal{M}_t \rightarrow \mathbb{R}$ modifies the base score as shown in Eq. \eqref{temporalenhancement}:

{\scriptsize
\begin{align}
I_t &= \mathcal{B}_{AoI}(\eta_{\text{temporal}}) + \Psi_{\text{focus}}(SF_s, s) + \mathcal{B}_{\text{sustained}}(|E_{\text{sustained}}|) \notag \\
&\quad + \mathcal{B}_{\text{duration}}(\mu_{\text{engagement}}) - \mathcal{P}_{\text{excess}}(|E|) \label{temporalenhancement}
\end{align}
}

where bonus functions $\mathcal{B}_{\cdot}$ and adjustment function $\mathcal{P}_{\cdot}$ are defined as Eq. \eqref{bonusa},\eqref{bonusb}, \eqref{bonusc}, \eqref{bonusd}:
{\scriptsize
\begin{align}
\mathcal{B}_{AoI}(\eta) &= 
\begin{cases}\label{bonusa}
\min(6, (\eta \times 100 - 30) \times 0.15) & \eta > 0.3 \\
-\min(4, (10 - \eta \times 100) \times 0.2) & \eta < 0.1 \\
0 & \text{otherwise}
\end{cases} \\\label{bonusb}
\mathcal{B}_{sustained}(n) &= \min(4, n \times 1.5) \\
\mathcal{B}_{duration}(\mu) &= 
\begin{cases}
\min(3, (\mu - 8) \times 0.2) & \mu > 8 \\
-\min(3, (3 - \mu) \times 0.8) & \mu < 3 \\
0 & \text{otherwise}
\end{cases} \label{bonusc}\\
\mathcal{P}_{excess}(n) &= \begin{cases}
\min(4, (n - 8) \times 0.6) & n > 8 \\
0 & \text{otherwise}
\end{cases}
\label{bonusd}
\end{align}
}

\subsubsection{Final Score Computation and Validation}
The final performance score combines spatial and temporal components as shown in Eq. \eqref{finalscore}:

\begin{equation}\label{finalscore}
    F_s = \max(0, \min(100, S_{base}(s) + \lambda_s \cdot I_t))
\end{equation}

Where $\lambda_s$ is a level-dependent temporal multiplier as shown in Eq. \eqref{multiplier}:

\begin{align}
\lambda_s = \begin{cases}
1.0 & S_{base} < 50 \\
0.9 & 50 \leq S_{base} < 70 \\
0.7 & 70 \leq S_{base} < 85 \\
0.4 & S_{base} \geq 85
\end{cases}
\label{multiplier}
\end{align}

Model performance is assessed using multiple validation metrics against ground truth game performance: MAE, RMSE, Pearson correlation, and Spearman correlation \cite{cohen2013applied} Eq. \eqref{metrixa},\eqref{metrixb}, \eqref{metrixc}, and \eqref{metrixd}:
{\scriptsize
\begin{align} 
\label{metrixa}MAE &= \frac{1}{n}\sum_{i=1}^{n}|F_i - A_{game,i}| \\
\label{metrixb}RMSE &= \sqrt{\frac{1}{n}\sum_{i=1}^{n}(F_i - A_{game,i})^2} \\
\label{metrixc}r_{Pearson} &= \frac{\sum_{i=1}^{n}(F_i - \bar{F})(A_{game,i} - \bar{A_{game}})}{\sqrt{\sum_{i=1}^{n}(F_i - \bar{F})^2}\sqrt{\sum_{i=1}^{n}(A_{game,i} - \bar{A_{game}})^2}} \\ 
\label{metrixd} \rho_{Spearman} &= 1 - \frac{6\sum_{i=1}^{n}d_i^2}{n(n^2-1)}
\end{align}
}

where $d_i$ represents the rank difference between $F_i$ and $A_{game,i}$.

\subsubsection{Model Constraints and Bounds}
We use multiple constraints to operate our mathematical model as shown in  Eq. \eqref{constraintsa}, \eqref{constraintsb}, \eqref{constraintsc}, \eqref{constraintsd}:

{\scriptsize
\begin{align}
\label{constraintsa} 0 \leq F_s &\leq 100, && \forall s \in \{1,2,3\} \\
\label{constraintsb} \tau_{\min} &= 400\,\text{ms} \leq \tau_{\text{sustained}} = 2500\,\text{ms} \\
\label{constraintsc} \sum_{i=1}^{4} P(Q_i) &= 1, && \text{(Prob. over quadrants)} \\
\label{constraintsd} |I_t| &\leq \text{max\_impact}_s, && \text{(Temporal impact bound)}
\end{align}
}

The proposed model rewards focused attention on stimuli, efficient navigation patterns, and sustained engagement while adjusting for distraction patterns, inefficient behavior, and inadequate stimulus focus. Level-adaptive thresholds ensure that difficulty progression is appropriate, from basic engagement (level 1) to advanced sustained attention (level 3).

\section{Experimental Findings}\label{experimentalfinding}
This section presents the findings and recommendations generated by our proposed scoring model. The model's performance is assessed using multiple validation metrics against the ground truth game performance.

TABLE \ref{tab:final_scores} summarizes the student's final performance across three levels. Each level includes a numerical score and a corresponding qualitative assessment, such as "Excellent" or "Good." It can be seen that students got promising scores in almost all levels. The student scored excellently in level 1 (97.7) and level 2 (99), with a slight dip to "Good" in level 3. The overall average of 95.0 out of 100 indicates outstanding performance.

\begin{table}[!ht]
\centering
\caption{Level-wise Final Scores and Assessments}
\label{tab:final_scores}
\begin{tabular}{|c|c|c|}
\hline
\textbf{Level} & \textbf{Final Score} & \textbf{Assessment} \\
\hline
1 & 99.7 & Excellent \\
\hline
2 & 99 & Excellent \\
\hline
3 & 98 & Excellent \\
\hline
\end{tabular}
\end{table}

TABLE \ref{tab:key_metrics} presents visual and interaction-based metrics: "Transitions" (number of gaze shifts), "AoI Focus\%" (attention on important screen areas), and "Objects" (number of interactive elements). These help interpret cognitive effort and complexity per level. The key metrics indicate that as the levels became more challenging, the student increased their eye movements (transitions rose from 105 to 107) but spent less focused time on critical areas (AoI Focus decreased from 40.9\% to 29.6\%), while interacting with more objects (111 to 101).

\begin{table}[!ht]
\centering
\caption{Attention and Complexity Metrics per level}
\label{tab:key_metrics}
\begin{tabular}{|c|c|c|c|}
\hline
\textbf{Level} & \textbf{Transitions} & \textbf{AoI Focus (\%)} & \textbf{Objects} \\
\hline
1&105& 40.9& 111\\\hline
2&107& 29.6& 101\\\hline
3&92&16.0& 103 \\
\hline
\end{tabular}
\end{table}

In TABLE \ref{tab:game_vs_model}, we compare the actual game accuracy with the scores generated by our model. The "Difference" column indicates the score gap between the two. The average game accuracy is 93.2\%, while our proposed model achieved a score of 99.2\%. Despite some fluctuations, the model's performance remains consistent, resulting in an overall gap of just 6.1\%.

\begin{table}[!ht]
\centering
\caption{Comparison of Game Accuracy vs Model Scores}
\label{tab:game_vs_model}
\scalebox{0.8}{
\begin{tabular}{|c|c|c|c|c|}
\hline
\textbf{level} & \textbf{Game Accuracy(\%)} & \textbf{Model Score} & \textbf{Difference} & \textbf{Assessment} \\
\hline
1 & 94.4 & 99.7 & 5.2 & Good\\\hline
2 & 97.8 & 99.0 & 1.3 & Excellent\\\hline
3 & 87.2 & 98.9 & 11.6 & Fair\\\hline
\textbf{Overall} & \textbf{93.2} & \textbf{99.2} & \textbf{6.1} & \\
\hline
\end{tabular}}
\end{table}

TABLE \ref{tabInteraction} includes raw gameplay statistics like mouse clicks, answers, and total events. Accuracy percentages are derived from these figures. It gives more profound insight into user interaction quality. It can be observed that the student correctly clicked the mouse 15 times out of 18 in the first level, 11 out of 11 in the second, and 10 out of 13 in the third level. Similarly, performance remained consistently high across all levels (94\% to approximately 98\%), indicating that the student maintained good game performance even as the difficulty increased. 

\begin{table}[!ht]
\centering
\caption{Detailed Interaction and Accuracy Metrics}
\label{tabInteraction}
\scalebox{0.8}{
\begin{tabular}{|c|c|c|c|c|}
\hline
\textbf{level} & \textbf{Mouse Clicks} & \textbf{Answers} & \textbf{Total Events} & \textbf{Accuracy (\%)} \\
\hline
1 & 15/18 & 36/36 & 54 & 94.4\\\hline
2 & 11/11 & 33/34 & 45 & 97.8\\\hline
3 & 10/13 & 31/34 & 47 & 87.2\\\hline
\end{tabular}}
\end{table}

TABLE \ref{tabvalidation} contains standard model evaluation metrics. Correlation values (Pearson and Spearman) test how well predicted scores follow actual performance. MAE and RMSE quantify prediction errors. The Pearson correlation of 0.389 indicates that eye-tracking scores have a substantial predictive value for game performance. The Mean Absolute Error of 6.05\% suggests that the framework's predictions are typically off by less than 6\%, which is quite accurate. 

\begin{table}[!ht]
\centering
\caption{Model Validation Statistics}
\label{tabvalidation}
\begin{tabular}{|l|c|}
\hline
\textbf{Metric} & \textbf{Value} \\
\hline
Pearson Correlation (r) & 0.389\\
\hline
Spearman Correlation ($\rho$) & 0.500\\
\hline
Mean Absolute Error (MAE) & 6.05\% \\
\hline
Root Mean Square Error (RMSE) & 7.41\% \\
\hline
\end{tabular}
\end{table}

\begin{figure*}[!ht]
 \centering
 \includegraphics[trim={0 0 0 3.1cm},clip,width=\textwidth]{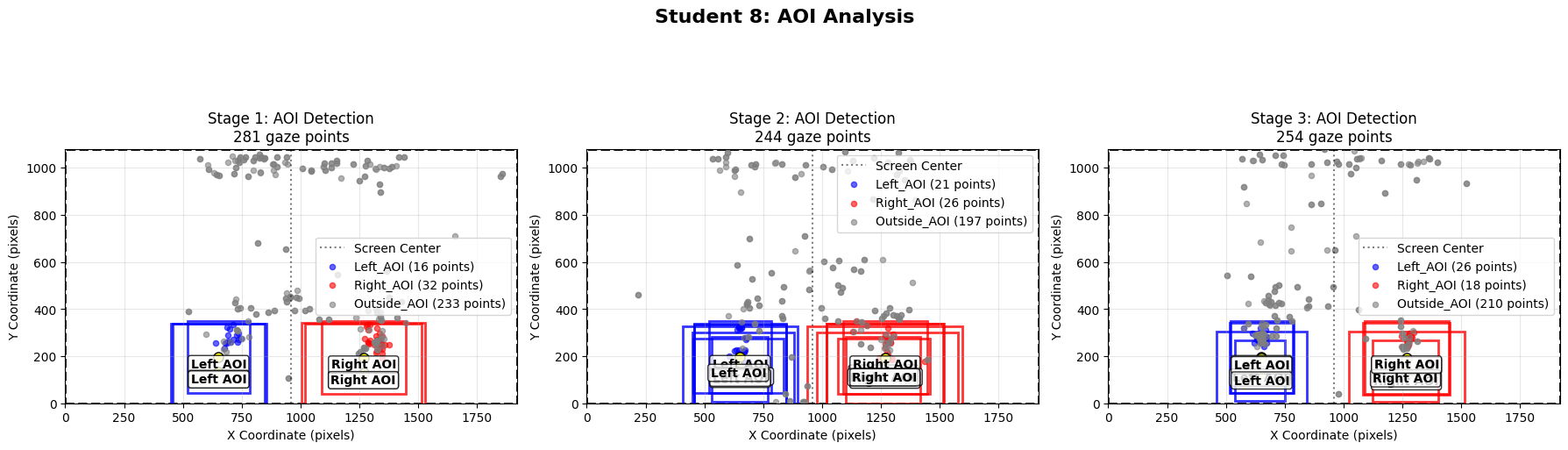}
 \caption{Area of Interest (AoI) detection across three levels.}
 \label{fig:AoI_detection}
\end{figure*}

FIGURE \ref{fig:AoI_detection} illustrates the framework's capabilities in detecting Areas of Interest (AoIs) by accurately classifying student gaze points in dynamically positioned regions across all three levels. 

In Stage 1, the proportion of gaze points inside the AoI was 17.1\%, while 82.9\% were outside. The engagement based on time spent in the AoI was recorded at 40.9\%. The viewing times in each quadrant were 28,586 ms for Q1, 29,096 ms for Q2, 49,690 ms for Q3, and 89,547 ms for Q4. Summing the times for Q3 and Q4 gives a total of 139,237 ms, which accounts for 70.7\% of the overall viewing time of 196,919 ms. In Stage 2, the inside-AOI viewing rate increased to 19.3\%, with 80.7\% of views being outside the AoI. The time-based engagement in this stage was 29.6\%. Quadrant viewing times were 30,855 ms for Q1, 20,597 ms for Q2, 52,274 ms for Q3, and 58,268 ms for Q4. Together, the times for Q3 and Q4 totaled 110,542 ms, representing 68.2\% of the total viewing time of 161,994 ms. In Stage 3, 17.3\% of gaze points were recorded inside the AoI, while 82.7\% were outside, with a time-based engagement of 16.0\%. The viewing times for this stage were 20,835 ms for Q1, 62,012 ms for Q2, 64,887 ms for Q3, and 26,850 ms for Q4. The combined time for Q3 and Q4 was 91,737 ms, which accounts for 52.5\% of the total viewing time of 174,584 ms. The visualization indicates that a majority of gaze points fall outside the defined AoI boundaries, which is expected, as students tend to scan the entire interface. In contrast, the designated AoI regions capture focused attention, which is crucial for learning assessment.
 
\begin{figure}[!ht]
 \centering
 \includegraphics[trim={0 0 0 1.1cm},clip,width=\columnwidth]{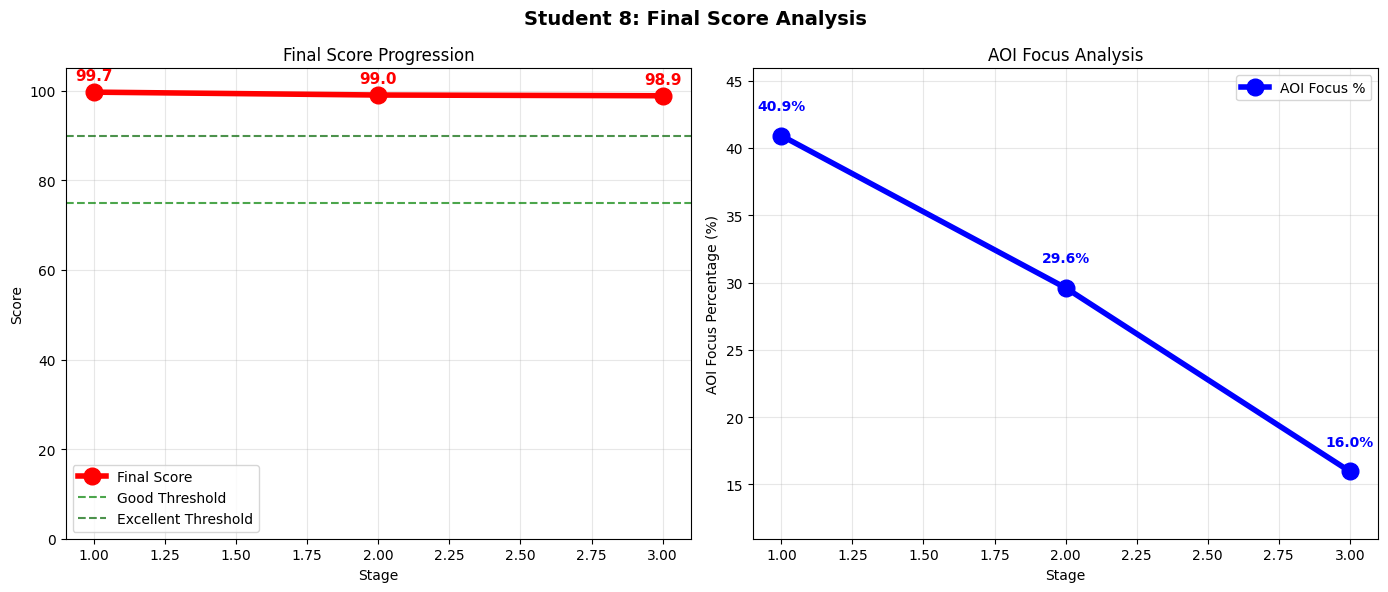}
 \caption{Student performance analysis}
 \label{figstudent_performance}
\end{figure}

FIGURE \ref{figstudent_performance} demonstrates the eye-tracking scoring framework's ability to capture nuanced learning behaviors across increasing difficulty levels. The left graph shows that this student maintains excellent performance scores (above 99) throughout all three levels, indicating strong adaptive learning capabilities and consistent attention control. Interestingly, the right panel reveals a systematic decrease in AoI focus from 40.9\% in level 1 to 16\% in level 3, which initially might seem contradictory to the high scores but reflects sophisticated learning behavior as students become more competent with the material; they require less concentrated focus on specific areas and can process information more efficiently through broader, strategic scanning patterns rather than intensive focal attention.

\begin{figure*}[!ht]
    \centering
    \includegraphics[trim={0 0 0 1cm},clip,width=0.9\textwidth]{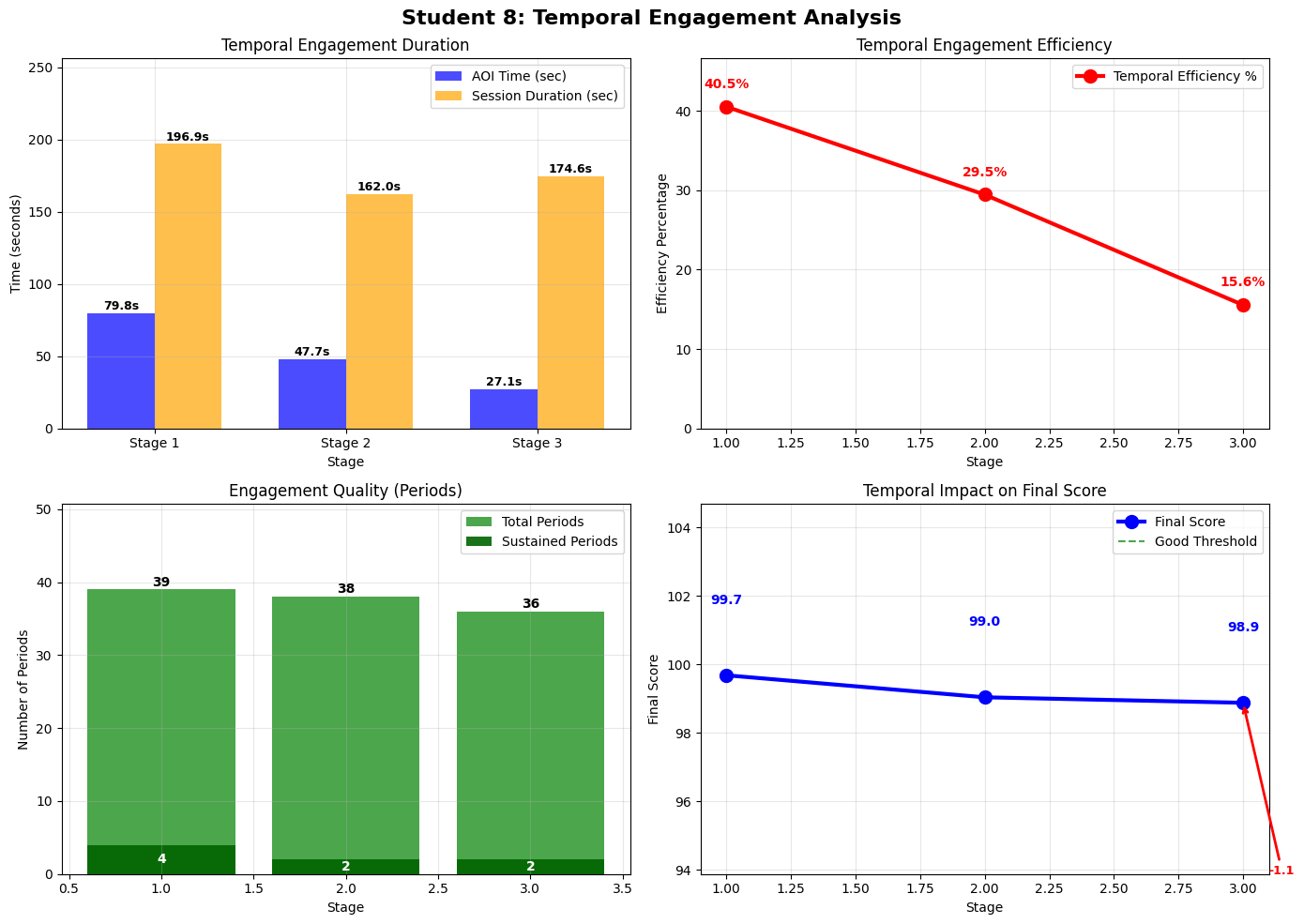}
 \caption{Comprehensive temporal engagement analysis for Student 10 across three difficulty levels.}
 \label{fig:temporal_engagement}
\end{figure*}

FIGURE \ref{fig:temporal_engagement} provides a detailed temporal analysis revealing sophisticated learning adaptation patterns across increasing difficulty levels. The top-left panel shows a systematic decrease in both AoI engagement time (from 79.4 seconds in level 1 to 27.1 seconds in level 3) and overall session duration, indicating more efficient task completion as expertise develops. The temporal efficiency (top-right) correspondingly decreases from 40.5\% to 15.6\%, which initially appears concerning but reflects a shift from intensive, focused attention to more strategic, distributed processing as students become more competent with the learning objects. The engagement quality analysis (bottom-left) demonstrates consistently high-quality attention patterns, with sustained engagement periods (dark green bars) representing the majority of total engagement across all levels, indicating deep, meaningful interaction rather than superficial scanning. It can be observed that as the level of difficulty increases, the duration of sustained attention periods decreases. Most notably, the temporal impact analysis (bottom-right) demonstrates how the framework adapts scoring based on task efficiency: Levels 1 and 2 remain unaffected by suboptimal timing, whereas Level 3 receives a minor negative adjustment (–1.1) due to reduced temporal efficiency, recognizing that the reduced AoI focus represents mastery-level efficiency rather than disengagement, ultimately contributing to the excellent final score of 98.9 points.

The proposed scoring model demonstrates exceptional performance and validation quality based on comprehensive testing and analysis. The framework achieved a final score correlation of r = 0.389 between eye-tracking-derived scores and actual game performance accuracy, indicating a significant relationship. Still, it is not strong enough to confidently predict individual student performance based solely on eye-tracking data.
The average final score across all analyzed sessions was 99.2 out of 100, reflecting excellent attention control and learning engagement from the test student. This high average score suggests that the scoring algorithm effectively captures superior learning behaviors while maintaining appropriate difficulty standards. This confirms that the eye-tracking scoring framework reliably translates complex attention patterns into meaningful educational metrics, making it suitable for high-stakes learning analytics and student performance evaluation in digital learning environments.

\section{Conclusion}\label{conclusion}
This paper focused on handling the limitations of traditional correct/incorrect scoring frameworks, which only indicate whether students provided the correct answer but offer no insight into how they learned or why they struggled. This paper presented a novel adaptive scoring framework that integrates several metrics, including spatial attention patterns, temporal engagement, and game performance data, to provide a holistic learning assessment beyond traditional game scoring. It further follows a progressive difficulty adaptation method with level-specific stimulus focus targets and dynamic weight adjustments that account for increasing cognitive load and learning complexity, and integrates temporal analysis capabilities, including engagement period detection, sustained attention rewards, and adaptive multiplier frameworks based on performance levels. Furthermore, it implemented an adaptive temporal impact framework that provides performance-scaled adjustments, preventing over-rewarding of high performers while maximizing improvement potential for struggling students. It established a multi-metric validation framework using MAE, RMSE, Pearson, and Spearman correlations with defined quality thresholds for deployment readiness assessment in educational settings. Finally, it bridged the gap between technical eye-tracking metrics and educational insights by providing explicit mappings between attention patterns and learning behaviors, enabling actionable pedagogical interventions.

\section*{Data availability}
Data will be made available on request.

\section*{Declaration of competing interest}
The authors share no conflict of interest.

\subsection{Acknowledgment}\label{ack}
The research leading to these results is within the frame of the ”EMPOWER. Design and evaluation of technological support tools to empower stakeholders in digital education” project, which has received funding from the European Union’s Horizon Europe programme under grant agreement No 101060918. Views and opinions expressed are, however, those of the author(s) only and do not necessarily reflect those of the European Union. Neither the European Union nor the granting authority can be held responsible for them.

\bibliographystyle{ieeetr}
\bibliography{ref-scoring}

\begin{thebibliography}{10}

\bibitem{edition1980diagnostic}
F.~EDITION, ``Diagnostic and statistical manual of mental disorders,'' {\em American Psychiatric Association, Washington, DC}, pp.~205--224, 1980.

\bibitem{von2022neurodevelopmental}
A.~von Gontard, J.~Hussong, S.~S. Yang, J.~Chase, I.~Franco, and A.~Wright, ``Neurodevelopmental disorders and incontinence in children and adolescents: Attention-deficit/hyperactivity disorder, autism spectrum disorder, and intellectual disability—a consensus document of the international children's continence society,'' {\em Neurourology and urodynamics}, vol.~41, no.~1, pp.~102--114, 2022.

\bibitem{skaramagkas2021review}
V.~Skaramagkas, G.~Giannakakis, E.~Ktistakis, D.~Manousos, I.~Karatzanis, N.~S. Tachos, E.~Tripoliti, K.~Marias, D.~I. Fotiadis, and M.~Tsiknakis, ``Review of eye tracking metrics involved in emotional and cognitive processes,'' {\em IEEE Reviews in Biomedical Engineering}, vol.~16, pp.~260--277, 2021.

\bibitem{coskun2022systematic}
A.~Coskun and K.~Cagiltay, ``A systematic review of eye-tracking-based research on animated multimedia learning,'' {\em Journal of Computer Assisted Learning}, vol.~38, no.~2, pp.~581--598, 2022.

\bibitem{sharma2022student}
P.~Sharma, S.~Joshi, S.~Gautam, S.~Maharjan, S.~R. Khanal, M.~C. Reis, J.~Barroso, and V.~M. de~Jesus~Filipe, ``Student engagement detection using emotion analysis, eye tracking and head movement with machine learning,'' in {\em International Conference on Technology and Innovation in Learning, Teaching and Education}, pp.~52--68, Springer, 2022.

\bibitem{garcia2019novel}
A.~Garc{\'\i}a-Baos, D.~Tomas, I.~Oliveira, P.~Collins, C.~Echevarria, L.~P. Zapata, E.~Liddle, H.~Super, {\em et~al.}, ``Novel interactive eye-tracking game for training attention in children with attention-deficit/hyperactivity disorder,'' {\em The primary care companion for CNS disorders}, vol.~21, no.~4, p.~26348, 2019.

\bibitem{merzon2022eye}
L.~Merzon, K.~Pettersson, E.~T. Aronen, H.~Huhdanp{\"a}{\"a}, E.~Seesj{\"a}rvi, L.~Henriksson, W.~J. MacInnes, M.~Mannerkoski, E.~Macaluso, and J.~Salmi, ``Eye movement behavior in a real-world virtual reality task reveals adhd in children,'' {\em Scientific reports}, vol.~12, no.~1, p.~20308, 2022.

\bibitem{aminoleslami2023classification}
S.~Aminoleslami, K.~Maghooli, N.~Sammaknejad, S.~Haghipour, and V.~Sadeghi-Firoozabadi, ``Classification of autistic and normal children using analysis of eye-tracking data from computer games,'' {\em Signal, Image and Video Processing}, vol.~17, no.~8, pp.~4357--4365, 2023.

\bibitem{oliveira2021computer}
J.~S. Oliveira, F.~O. Franco, M.~C. Revers, A.~F. Silva, J.~Portolese, H.~Brentani, A.~Machado-Lima, and F.~L. Nunes, ``Computer-aided autism diagnosis based on visual attention models using eye tracking,'' {\em Scientific reports}, vol.~11, no.~1, p.~10131, 2021.

\bibitem{lee2023use}
D.~Y. Lee, Y.~Shin, R.~W. Park, S.-M. Cho, S.~Han, C.~Yoon, J.~Choo, J.~M. Shim, K.~Kim, S.-W. Jeon, {\em et~al.}, ``Use of eye tracking to improve the identification of attention-deficit/hyperactivity disorder in children,'' {\em Scientific Reports}, vol.~13, no.~1, p.~14469, 2023.

\bibitem{costescu2023mushroom}
C.~Costescu, C.~David, A.~Roșan, P.~Ferreira, A.~Ferreira, L.~Vera, and G.~Herrera, ``Mushroom hunters: A digital game for assessing and training sustained attention in children with neurodevelopmental disorders,'' in {\em Methodologies and Intelligent Systems for Technology Enhanced Learning, Workshops - 13th International Conference} (Z.~Kubincov{\'a}, F.~Caruso, T.-e. Kim, M.~Ivanova, L.~Lancia, and M.~A. Pellegrino, eds.), (Cham), pp.~78--86, Springer Nature Switzerland, 2023.

\bibitem{CCPT}
J.~F.~S. Ph.D., M.~S.~T. M.Sc., D.~I. S.-W. M.Sc., P.~Jan K. Buitelaar~M.D., H.~S.-B. Ph.D., F.~C.~V. M.D., T.~C.~P. M.Sc., and D.~I. B.~P. and, ``Sustained attention and executive functioning performance in attention-deficit/hyperactivity disorder,'' {\em Child Neuropsychology}, vol.~11, no.~3, pp.~285--294, 2005.
\newblock PMID: 16036452.

\bibitem{projectempowerProjectEMPOWER}
HorizonEurope, ``Project empower, design and evaluation of technological support tools to empower stakeholders in digital education.'' \url{https://project-empower.eu/}, 2024.
\newblock [Accessed 08-05-2025].

\bibitem{holmqvist2011eye}
K.~Holmqvist, M.~Nystr{\"o}m, R.~Andersson, R.~Dewhurst, H.~Jarodzka, and J.~Van~de Weijer, {\em Eye tracking: A comprehensive guide to methods and measures}.
\newblock oup Oxford, 2011.

\bibitem{duchowski2017eye}
A.~T. Duchowski and A.~T. Duchowski, {\em Eye tracking methodology: Theory and practice}.
\newblock Springer, 2017.

\bibitem{rayner2009}
K.~Rayner, ``Eye movements and attention in reading, scene perception, and visual search,'' {\em Quarterly journal of experimental psychology}, vol.~62, no.~8, pp.~1457--1506, 2009.

\bibitem{henderson2003human}
J.~M. Henderson, ``Human gaze control during real-world scene perception,'' {\em Trends in cognitive sciences}, vol.~7, no.~11, pp.~498--504, 2003.

\bibitem{posner2011attention}
M.~I. Posner, {\em Attention in a social world}.
\newblock Oxford University Press, 2011.

\bibitem{fan2002testing}
J.~Fan, B.~D. McCandliss, T.~Sommer, A.~Raz, and M.~I. Posner, ``Testing the efficiency and independence of attentional networks,'' {\em Journal of Cognitive Neuroscience}, vol.~14, no.~3, pp.~340--347, 2002.

\bibitem{maclean2010intensive}
K.~A. MacLean, E.~Ferrer, S.~R. Aichele, D.~A. Bridwell, A.~P. Zanesco, T.~L. Jacobs, B.~G. King, E.~L. Rosenberg, B.~K. Sahdra, P.~R. Shaver, {\em et~al.}, ``Intensive meditation training improves perceptual discrimination and sustained attention,'' {\em Psychological science}, vol.~21, no.~6, pp.~829--839, 2010.

\bibitem{just1976eye}
M.~A. Just and P.~A. Carpenter, ``Eye fixations and cognitive processes,'' {\em Cognitive psychology}, vol.~8, no.~4, pp.~441--480, 1976.

\bibitem{paas2016cognitive}
F.~Paas, J.~E. Tuovinen, H.~Tabbers, and P.~W. Van~Gerven, ``Cognitive load measurement as a means to advance cognitive load theory,'' in {\em Cognitive load theory}, pp.~63--71, Routledge, 2016.

\bibitem{sweller1998cognitive}
J.~Sweller, J.~J. Van~Merrienboer, and F.~G. Paas, ``Cognitive architecture and instructional design,'' {\em Educational psychology review}, vol.~10, pp.~251--296, 1998.

\bibitem{cohen2013applied}
J.~Cohen, P.~Cohen, S.~G. West, and L.~S. Aiken, {\em Applied multiple regression/correlation analysis for the behavioral sciences}.
\newblock Routledge, 2013.

\end{thebibliography}

\begin{IEEEbiography}{Abdul Rehman} is a Ph.D. candidate in Computer Science at the Department of Computer Science, Electrical Engineering and Mathematical Sciences, Western Norway University of Applied Sciences, Bergen, Norway. His research interests include privacy-preserving machine learning, eye tracking, human-computer interaction, and educational technologies.
\end{IEEEbiography}

\begin{IEEEbiography}{Ilona Heldal} is a Professor at the Department of Computer Science, Electrical Engineering and Mathematical Sciences, Western Norway University of Applied Sciences, Bergen, Norway. She holds a Ph.D. in technology and has extensive experience in virtual reality, human-computer interaction, and collaborative learning environments.
\end{IEEEbiography}

\begin{IEEEbiography}{Cristina Costescu} is an Associate Professor in the Special Education Department at Babeș-Bolyai University in Cluj-Napoca, Romania. Her research focuses on neurodevelopmental disorders, digital education, and executive functions. She has been involved in several projects and grants, including investigating executive functions and emotion regulation strategies in atypical development and using robot-enhanced therapy to improve socio-emotional skills in children with Autism Spectrum Disorder (ASD). She has also co-authored publications on topics such as the effectiveness of digitally mediated social stories and the use of technology to support math education for children with intellectual disabilities.
\end{IEEEbiography}

\begin{IEEEbiography}{Carmen David} is affiliated with the Special Education Department at Babeș-Bolyai University, Cluj-Napoca, Romania. Her work includes research on social communication predictors in autism spectrum disorder and the relationship between cognitive and emotional processes in children and adolescents with neurodevelopmental disorders. She has co-authored publications with Cristina Costescu and other researchers in the field.
\end{IEEEbiography}

\begin{IEEEbiography}{Jerry Chun-Wei Lin} is a Professor at the Department of Computer Science, Electrical Engineering and Mathematical Sciences, Western Norway University of Applied Sciences, Bergen, Norway. His research interest includes but bit limited to data mining, deep learning, and privacy-preserving algorithms. 
\end{IEEEbiography}

\EOD
\end{document}